\documentclass[letterpaper]{article}
\newcommand{\affil}[1]{$^{\rm #1}$}
\usepackage[authoryear]{natbib}
\bibpunct{(}{)}{;}{a}{}{,}
\usepackage{graphicx}
\date{} 
\newcommand{\kms}{\mbox{km\,s$^{-1}$}}
\newcommand{\dm}{$\Delta m_{15}(B)$}
\newcommand{\snia}{SN~Ia}
\newcommand{\sneia}{SNe~Ia}
\def\ion#1#2{#1$\;${\small\rm{#2}}\relax}
\title{\large\bf\flushleft Near-Infrared Properties of Type~Ia Supernovae}
\author{\parbox{\textwidth}{\flushleft
\vspace{-0.5cm}
{\it M. M. Phillips\affil{A,B}}\\
\vspace{0.4cm}
{\small \affil{A}\,Las Campanas Observatory, Carnegie Observatories, Casilla 601, La Serena, Chile}\\
{\small \affil{B}\,Email: mmp@lco.cl}}}
\begin{document}
%
\begin{changemargin}{.8cm}{.5cm}
\begin{minipage}{.9\textwidth}
\vspace{-1cm}
\maketitle
%
%
\small{\bf Abstract:}

The photometric properties of Type~Ia supernovae (\sneia) in the near-infrared as garnered from observations made over the last 30 years are reviewed.  During this period, light curves for more than 120 nearby \sneia\ have been published, revealing considerable homogeneity  but also some fascinating differences.  These data have confirmed that, for all but the fastest declining objects, \sneia\ are essentially perfect standard candles in the near-infrared, displaying only a slight dependence of peak luminosity on decline rate and color.


\medskip
\medskip
\end{minipage}
\end{changemargin}

\small

\section{Introduction}
\label{sec:intro}

Supernovae (SNe) were first grouped into two separate classes by \cite{minkowski41} on the basis of the presence or absence of hydrogen lines in the spectrum.  The majority of the SNe which lack hydrogen belong to a fascinating subclass of exploding stars now called ``\sneia''.  More than 50 years ago, it was understood that the high degree of homogeneity displayed by \sneia\ is due to the fact that they are the end result of the total thermonuclear disruption of a white dwarf in a close binary system \citep{hoyle60}.  Nevertheless, our understanding of the nature of the progenitors and the details of the explosion mechanism(s) remains frustratingly sketchy.  In spite of this, \sneia\ have played a fundamental role in the revolution in observational cosmology that has taken place since the early-1990s.  Not only have observations of \sneia\ played a fundamental role in the determination of the Hubble constant to a precision of better than 5\% \citep{riess11}, but they also led to the serendipitous discovery of dark energy \citep{riess98,perlmutter99}. 

Investigations carried out since the late-1980s have demonstrated that \sneia\ are not perfectly homogenous.  Rather, their successful utility as cosmological distance indicators at optical wavelengths rests on the discovery of an empirical correlation between the absolute magnitude at maximum light and the decline rate (or width) of the light curve \citep{phillips93}.  The sense of this correlation is that slower-evolving \sneia\ are intrinsically more luminous than faster-evolving ones, with the effect being largest in the blue where the total observed range in peak luminosity is $\sim$2 mag \citep{hamuy96a}.  The absolute magnitudes of \sneia\ are also correlated with color as would be expected if their light were extinguished by dust.  However, the ratio of the total-to-selective extinction, $R_V$, derived from the most-highly reddened \sneia\ is often found to be unusually low ($R_V \sim$1.5--2.1) \citep[see][and references therein]{krisciunas07}, suggesting that the situation is more complex than simple dust reddening alone.

Until the first half of the 1980s, the near-infrared (NIR) properties of \sneia\ were largely unexplored.  In a landmark paper, \citet{elias81} presented $JHK$ photometry of the three nearby \sneia: 1980N and 1981D, both of which appeared in the Fornax cluster member NGC~1316 (Fornax~A), and 1981B, which was hosted by NGC~4536 in the Virgo cluster.  The light curves of all three SNe were remarkably similar and characterized by a distinctive double maximum\footnote{\citet{elias81} pointed out that published and unpublished $I$-band photometry of SN~1972E had shown a similar behavior.}.  Attention was also called to the complexity of the $J-H$ color evolution, which was interpreted as being due to the formation of a broad, transient absorption feature at $\sim$1.2~$\mu$m during the phase of the light curve leading up to the second maximum.  The dispersion in the absolute magnitudes of the three SNe was less than the uncertainties in the relative distances, prompting the authors to suggest that NIR light curves of \sneia\ might be useful for measuring distances within the local universe.  This important paper was followed a few years later by a larger study \citep{elias85} which included the first NIR Hubble diagram for \sneia\footnote{\citet{elias85} were also the first to use the nomenclature ``Ia'' and ``Ib'' to differentiate between Type~I SNe that we now understand to be of thermonuclear vs. stripped core-collapse origin.}.  A dispersion of only 0.1--0.2~mag was found for a sample of six unambiguously-identified \sneia, reemphasizing the potential utility of these objects as cosmological distance indicators.

In this paper, I review what has been learned about the NIR photometric properties of \sneia\ since the pivotal work of Elias and collaborators.  Much of the paper concentrates on the NIR light curves of normal \sneia, what they tell us about the dust extinction in the line of sight when combined with optical photometry, and their use for determining precise distances.  This is followed by a discussion of the nature of the fast-declining SN~1986G-like and SN~1991bg-like events as considered from the NIR perspective, and brief mention of the NIR properties of a few peculiar subgroups of \sneia.  I have taken advantage of this review to highlight recent results from the Carnegie Supernova Project \citep[CSP;][]{hamuy06}, some of which are as yet unpublished.  For this I beg the reader's indulgence.

\section{Near-Infrared Light Curves}
\label{sec:nirlcurves}

During the 17 years following the publication of \citet{elias81}, NIR photometry was obtained for only a few more \sneia.  The contrast in luminosity of a \snia\ with respect to its host galaxy is typically less in the NIR than at optical wavelengths, and measuring the brightness of a SN immersed in the light of its host using the instrumentation available during these years, which generally employed single-element detectors, was a significant challenge.  With one obvious exception, the data for the few \sneia\ observed were consistent with the light curves of the six \sneia\ presented by \citet{elias81,elias85}, adding to the impression of a high level of homogeneity.  The exception was SN~1986G, for which excellent photometry in $JHK$ was obtained by \citet{frogel87} beginning five days before the epoch of $B$ maximum.  SN~1986G appeared in the dust lane of NGC~5128 (Centauras~A) and was observed extensively at optical wavelengths, both photometrically and spectroscopically \citep{phillips87,cristiani92}.  The $B$ light curve declined from maximum much more rapidly than normal, and the spectroscopic evolution, while similar to that of slower-declining \sneia, displayed subtle differences at maximum light.  \citet{frogel87} showed that the NIR light curves of SN~1986G were also significantly different from those of normal \sneia.  In particular, the secondary maximum occurred $\sim$15 days earlier than it does for normal \sneia, and appeared more as a shoulder rather than a distinct peak.  In \S\ref{sec:fastsneia}, I will return to discuss these fast-declining events in more detail.

The first pre-maximum $JHK$ photometry of a normal \sneia\ was obtained for SN~1998bu by \citet{jha99} and \citet{hernandez00}.  These observations showed that maximum light in the NIR occurs approximately 5 days {\em before} the epoch of $B$ maximum.  Thus, in order to catch a \snia\ still on the rise in $JHK$, observations must begin typically 6--8 days before $B$ maximum.  This, along with the above-mentioned contrast issue between the brightness of the SN and its host, are the main drawbacks of observing \sneia\ in the NIR.

Between 2000 and 2004, the number of \sneia\ with NIR light curves tripled, owing in large part to a collaborative program of observations carried out at the Cerro Tololo Inter-American and Las Campanas Observatories in Chile \citep{krisciunas01,krisciunas03,krisciunas04b,krisciunas04c} that was made possible by the development and accessibility of panoramic NIR detectors.  With this larger sample and the improving quality and coverage of the observations, differences in light-curve morphology and colors now became evident even among normal \sneia.  While most of these differences correlate with the $B$-band decline rate, evidence was clearly seen of variations in light-curve shape --- particularly the strength of the secondary maximum --- for SNe with the {\em same} decline rates \citep{krisciunas01}.

In recent years, NIR photometry of \sneia\ has become relatively commonplace.  Two surveys in particular deserve special mention.  The first of these is being carried out using the 1.3~m PARITEL at the F.~L. Whipple Observatory at Mt. Hopkins, Arizona, and has resulted in the publication of $JHK_s$ light curves for 21 \sneia\ to date \citep{wood-vasey08}.  The second is the CSP, which employed the 1.0~m and 2.5~m telescopes at Las Campanas Observatory (LCO) between 2004-2009 to obtain $YJH$ light curves (plus a small amount of $K_s$ photometry) for $\sim$100 \sneia.  The NIR data for 70 of these have been published by \citet{contreras10} and \citet{stritzinger11}.  Extensive optical photometry was also obtained for both the PARITEL and CSP SNe, adding considerable extra value to these data sets.

\subsection{General Properties}
\label{sec:genprops}

\begin{figure}[h]
\begin{center}
\includegraphics[scale=.5, angle=0]{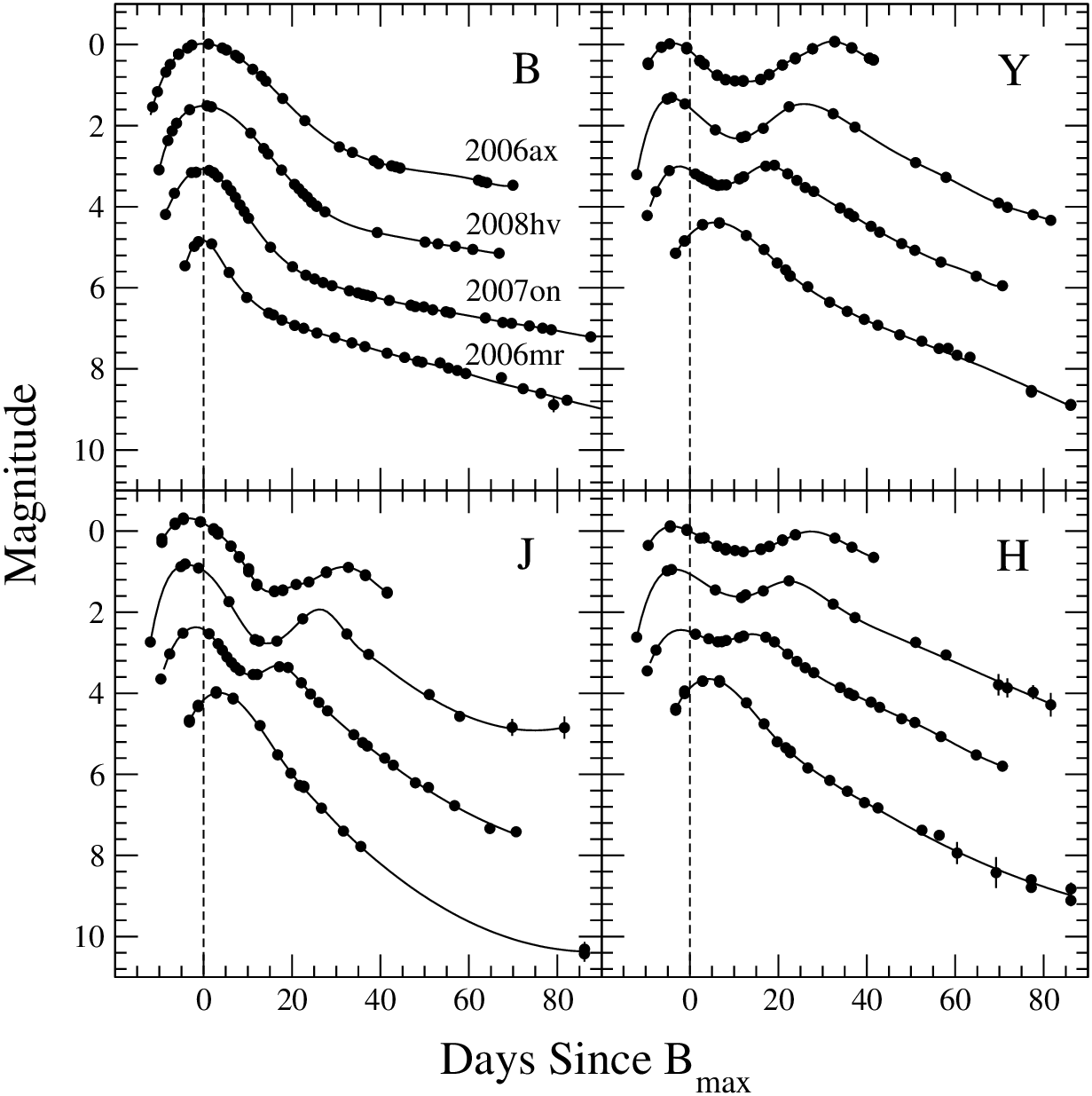}
\caption{CSP light curves in $BYJH$ of four \sneia\ covering a wide range of decline rates: SN~2006ax ($\Delta$m$_{15}$ = 0.97), SN~2008hv ($\Delta$m$_{15}$ = 1.25), SN~2007on ($\Delta$m$_{15}$ = 1.62), and SN~2006mr ($\Delta$m$_{15}$ = 1.71).  Arbitrary offsets in magnitude have been subtracted from  the data to facilitate comparison.}\label{fig:lcurves}
\end{center}
\end{figure}

Figure~\ref{fig:lcurves} shows the $BYJH$ light curves of four \sneia\ observed by the CSP that cover a range of decline rates.  A commonly-used parameter for quantifying the decline rate is \dm, which was defined by \citet{phillips93} as the amount in magnitudes that the $B$ light curve declines during the first 15 days after maximum light.  In this paper, I will also use the closely-related parameter $\Delta$m$_{15}$ employed by the SNooPy (SuperNovae in Object-Oriented Python) light-curve fitting package \citep{burns11}, which is derived from a fit to {\em all} the available photometry ($uBgVriYJH$ in the case of the CSP observations) and not just the $B$ light curve.  The four SNe plotted in Figure~\ref{fig:lcurves} have $\Delta$m$_{15}$ = 0.97 -- 1.71, which covers most of the range of observed decline rates of \sneia.  

As is apparent in Figure~\ref{fig:lcurves}, the differences in decline rate reflect differences in the overall width of the rise and fall from $B$-band maximum.  Indeed, this is the basis of the stretch method \citep{perlmutter97}, which is a commonly-used alternative method of quantifying the luminosity-decline rate relationship.  Also obvious in Figure~\ref{fig:lcurves} is the prominent secondary maximum observed in the NIR.  Although visible from the $I$-band through the $K$-band, and seen as a shoulder in $V$ and $R$, the secondary maximum reaches its maximum expression in $Y$ and $J$.  Interestingly, in both $Y$ and $H$, the secondary maximum is typically as bright as (and sometimes even brighter than) the primary maximum.   Figure~\ref{fig:lcurves} also illustrates that the time interval between the primary and secondary maxima decreases as a function of decline rate (or luminosity).  Note that for most \sneia, primary maximum in the NIR occurs several days {\em before} the epoch of $B$ maximum, but for some very fast-declining events (e.g., SN~2006mr in Figure~\ref{fig:lcurves}), it can occur {\em after} $B$ maximum (see also \S\ref{sec:fastsneia}).

The origin of the secondary maximum in \sneia\ has been treated by several authors. \citet{hoeflich95} argued that the secondary maximum was a temperature-radius effect produced in models where the photospheric radius continues to increase well after maximum.   \citet{pinto00} explained it as the release of trapped radiation due to a sudden decrease in the flux mean opacity when singly-ionized Fe becomes the dominant ion in the inner ejecta.  From detailed radiative transfer models, \citet{kasen06} confirmed that the secondary maximum is a direct consequence of the ionization evolution of the Fe-group elements and explored the dependence of its morphology on the physical properties of the ejecta.

Studies of the $I$-band light curves \citep{hamuy96b,krisciunas01,folatelli10} have shown that the strength of the secondary maximum in \sneia\ decreases as a function of decline rate (or luminosity), fading to invisibility in the very fastest-declining objects (see SN~2006mr in Figure~\ref{fig:lcurves}).  The models indicate that both the strength and timing of the secondary maximum are governed principally by the amount of radioactive $^{56}$Ni produced in the explosion \citep{hoeflich95,kasen06}.  Nevertheless, at a given decline rate, real differences are observed in the strength and morphology of the secondary maximum \citep{krisciunas01,folatelli10} which may be attributable to secondary parameters such as the mixing into the ejecta of the $^{56}$Ni \citep{kasen06}.  These differences in the strength of the secondary maximum can provide valuable insight into the physics of the explosion mechanism, but at the same time represent a significant problem for template fitting in the $IJHK$ bands by light curve fitters such as SNooPy, where the strength of the secondary maximum is assumed to be a smoothly varying function of decline rate \citep{burns11}. 

The nature of the striking depression at $\sim$1.2~$\mu$m that develops during the phase of the light curve leading up to the second maximum sparked considerable discussion after attention was drawn to it by \citet{elias81}\footnote{\citet{kirshner73} had been the first to detect this feature from optical and NIR observations of SN~1972E.}.  Although the rapidity with which the feature appeared and disappeared suggested to \citet{elias81} that it was produced by a single species, the correct explanation did not come until more than 10 years later when \citet{spyromilio94} suggested that the broad ``absorption'' was actually due to a lack of significant opacity at these wavelengths.  \citet{pinto00} went on to show that, due to a shift to more neutral species after maximum light, the photosphere recedes rapidly to the center of the SN in the NIR.  In effect, the NIR spectrum transforms into a broad-line emission nebula which happens to have no strong lines at 1.2~$\mu$m.  This explanation was observationally confirmed by \citet{hamuy02}, who obtained the first continuous sequence of NIR spectra of a \snia\ (1999ee).

Due to the observational difficulties, relatively little NIR photometry has been obtained of \sneia\ at epochs beyond three months past maximum.  \citet{elias83} managed to obtain $JHK$ measurements of SN~1981B at $\sim$4.5 months after maximum, $H$ observations of SN~1980N at $\sim$7.5 months past maximum, and detected SN~1981D in $J$ and $H$ $\sim$9 months after maximum.  The $H$ band data appeared to be consistent with a steady linear decline of $\sim$0.013~mag~day$^{-1}$ from day +100 onwards.  The \citet{elias83} observations were heroically obtained using a single-channel photometer, with substantial background corrections obtained by averaging measurements of the background host galaxy light near the SNe.

Fortunately, the advent of NIR arrays made late-epoch photometry considerably more feasible, and \citet{spyromilio04} were able to obtain solid detections of SN~1998bu in $H$ at epochs of +250 and +344.  These observations revealed that the SN flux had faded only slightly between the two epochs, and was considerably brighter than a decline rate of $\sim$0.013~mag~day$^{-1}$ would have predicted.  Around the same time, \citet{sollerman04} published $J$ and $H$ photometry of the spectroscopically-peculiar SN~2000cx at five epochs between +369 and +483 days, and a few years later \citet{stritzinger07} presented $JHK$ measurements of the normal SN~2001el obtained from +316 to +445 days. In both cases, the late-time NIR emission remained essentially constant.  Such a flattening of the light curve at late epochs was argued by \citet{sollerman04} to be consistent with theoretical predictions of the gradual transition of the emission from the optical to the NIR as the ejecta expand and the temperature decreases.  However, no evidence was seen of the ``infrared catastrophe'' (IRC) predicted by \citet{axelrod80}, where most of the emission shifts from optical/NIR wavelengths to the mid- and far-IR fine-structure lines when the ejecta temperature falls below a certain threshold.

To date, the only fast-declining \snia\ observed at very late epochs is SN~2003hv (\dm $\sim$1.6), for which ground- and space-based NIR photometry was acquired at phases from +344 to +786 days \citep{leloudas09}.  At these epochs, the evolution of the NIR flux was again nearly flat.  As was also observed for SNe~2000cx and 2001el, the relative contribution of the NIR bands to the total $U$-to-$K$ luminosity increased steadily from a low of $< 5$\% at day +150 to nearly 40\% by day +700.  Interestingly, the derived $^{56}$Ni mass at late epochs was found to be significantly-less than the value at maximum light given by Arnett's law \citep{arnett82}, implying that $> 40$\% of the total luminosity of the SN was being emitted in the mid- and far-infrared by day +500, suggesting that the IRC may actually have occurred in the densest portions of the ejecta by day +350.  The recent discovery of SN~2011fe in M101 \citep{nugent11}, which reached $V \sim 10$ at maximum light, will provide the first ever opportunity to follow a normal \snia\ in the NIR, both photometrically and spectroscopically, to epochs of +3 years and beyond.

\subsection{Dust Reddening}
\label{sec:dust}

One of the most vexing problems in dealing with the light curves of \sneia\ is determining the extinction of the SN light due to dust in the interstellar medium of its host galaxy and/or the circumstellar environment of the progenitor system itself.  \citet{krisciunas00} were the first to emphasize the advantage of combining optical and NIR photometry to estimate the visual extinction, $A_V$, to a \snia.  Since the extinction in the $H$ band is 5.5 times less than in $V$, the color excess $E(V-H) = A_V - A_H$ $\approx A_V$ for all but the most highly reddened objects.  The measurement of color excesses as a function of wavelength also provides invaluable information on the shape of the reddening curve.

\begin{figure}[h]
\begin{center}
\includegraphics[scale=.5, angle=0]{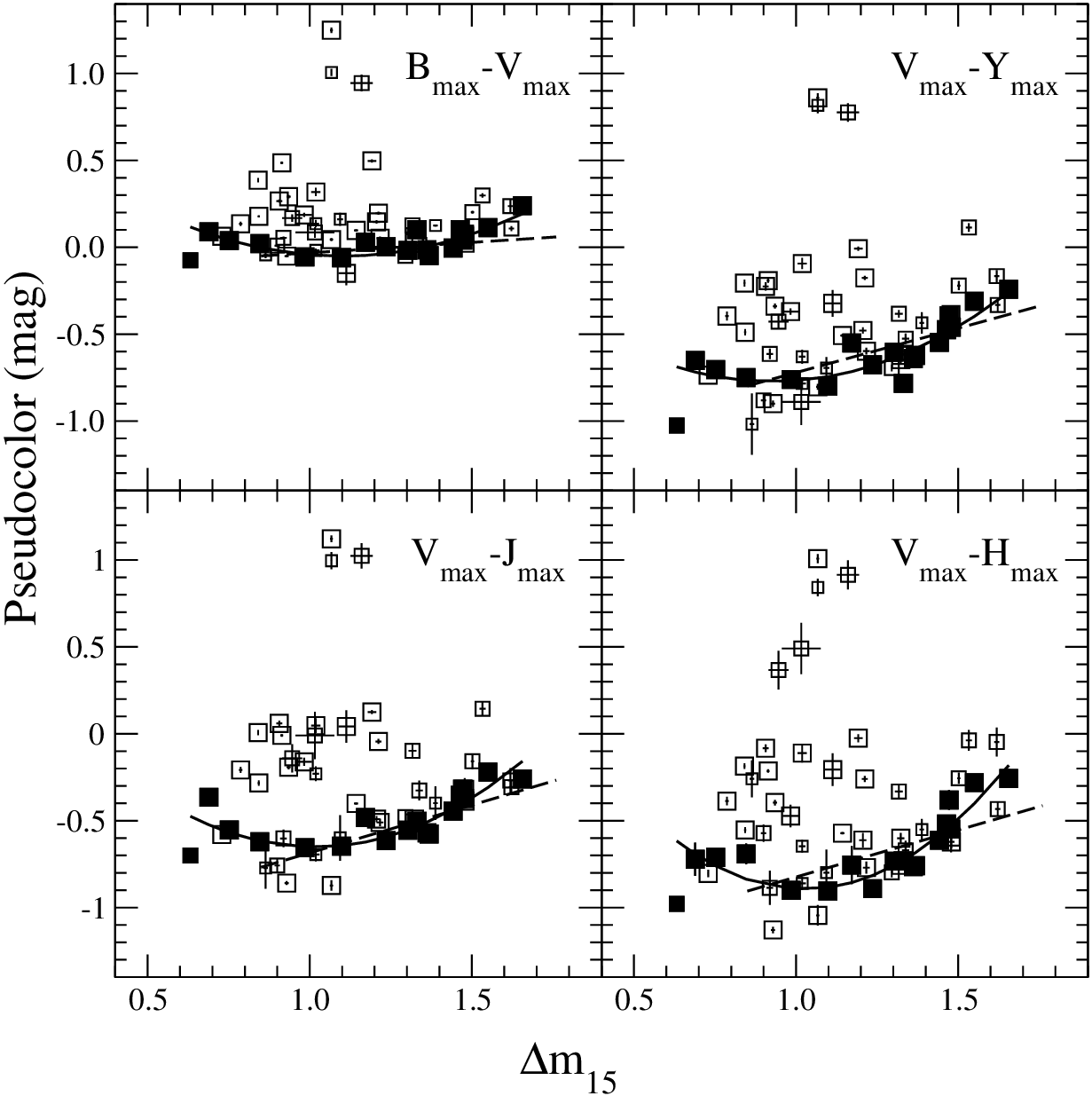}
\caption{Plots of pseudocolor vs. decline rate for 59 \sneia\ observed by the CSP.  The size of the points is related to the precision of the pseudocolor measurements. Solid symbols correspond to the low-reddening subsample (see text).  The dashed lines are the intrinsic pseudocolor vs. decline rate relations determined by \citet{folatelli10}; the continuous curves are second-order polynomial fits to the low-reddening subsample.}\label{fig:colors}
\end{center}
\end{figure}

Figure~\ref{fig:colors} displays pseudocolors plotted versus decline rate for 59 \sneia\ observed by the CSP.  The term ``pseudocolor'' is employed to stress the fact that these quantities do not represent the actual color of the SNe at any time but are just the difference between the magnitudes at maximum light of two bands, which occur at slightly different times.  A blue edge to the distribution of pseudocolors is clearly visible in each of the panels of Figure~\ref{fig:colors}.  Indicated as solid symbols in the figure is a subsample of SNe which appear to have suffered little dust extinction as determined using the Lira relationship \citep[for details, see][]{folatelli10}.  As expected, this low-reddening subsample lies along the blue edge.  Figure~\ref{fig:colors} displays the linear fits to the low-reddening SNe derived by \citet{folatelli10} from a smaller sample of CSP \sneia. The larger sample in Figure~\ref{fig:colors} suggests that the blue edge may actually be curved, and so second-order polynomial fits to the low-reddening subsample are also shown.

Figure~\ref{fig:colors} implies that the majority of \sneia\ do not suffer large amounts of dust extinction.  This is consistent with previous studies which, under the assumption that the distribution of host extinction\footnote{Here the term ``host extinction'' includes all sources of dust extinction (intergalactic, interstellar, circumstellar) produced between the Milky Way and the SN.} is well-described by an exponential, $P(A_V) \propto {\rm exp} ( -A_V / \tau_V )$, where $\tau_V \sim$ 0.3--0.5~mag \citep{jha07,kessler09}.

\begin{figure}[h]
\begin{center}
\includegraphics[scale=.5, angle=0]{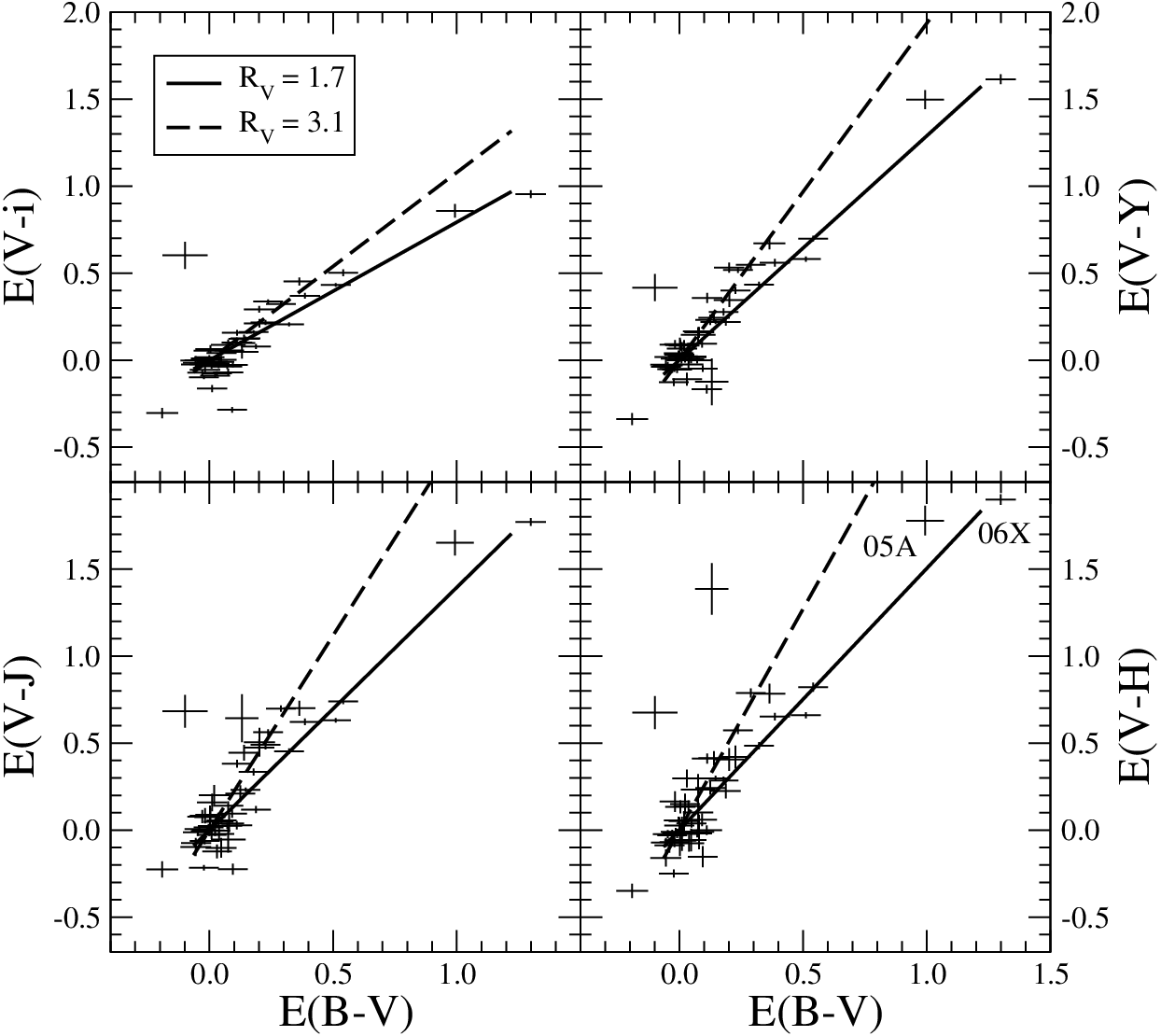}
\caption{Comparison of color excess estimates $E(V-X)$, for $X = iYJH$, with $E(B-V )$ for 52 CSP \sneia. The solid lines represent the slope which corresponds to $R_V = 1.7$; the dashed lines indicate the slope for the canonical Galactic value of $R_V = 3.1$.}\label{fig:color_excesses}
\end{center}
\end{figure}

Assuming the distribution of SN colors seen in Figure~\ref{fig:colors} is due to dust extinction, then the vertical offset of each SN from the blue edge provides a measurement of its color excess.  Figure~\ref{fig:color_excesses} shows the resulting ($V-NIR$) color excesses --- $E(V-i)$, $E(V-Y)$, $E(V-J)$, and $E(V-H)$ --- plotted versus $E(B-V)$.  The second-order polynomial fits shown in Figure~\ref{fig:colors} were used to define the blue edges.  The diagonal lines in each panel of Figure~\ref{fig:color_excesses} indicate the expected correlations of the color excesses for two different values of $R_V$ assuming the standard Galactic reddening law of \citet{cardelli89}.  The dashed line represents the canonical Galactic value of $R_V = 3.1$, and the solid line corresponds to $R_V = 1.7$.  

While there are a few SNe which lie a considerable distance from either reddening line, the vast majority fall close to the cone defined by the two values of $R_V$.  Interestingly, six of the eight most highly reddened SNe fall close to the $R_V = 1.7$ relation.  A detailed analysis by \citet{folatelli10} of all of the CSP photometry (including the important $u$-band) of two of these SNe, 2005A and 2006X, showed that their color excesses were well matched by the model of \cite{goobar08} where multiple scattering of photons by circumstellar dust steepens the effective extinction law.  Independent evidence for the existence of circumstellar material associated with SN~2006X was found by \citet{patat07}, who observed short-term variations in the interstellar \ion{Na}{I}~D absorption lines during its early-time spectral evolution.  Nevertheless, Figure~\ref{fig:color_excesses} indicates that the color excesses of the majority of the SNe with $E(B-V) < 0.3$ are consistent with the standard Galactic value of $R_V \sim 3$.  A similar result was found recently by \citet{folatelli10}, \citet{mandel11}, and \citet{chotard11}.

Figure~\ref{fig:color_excesses} suggests that the reddening of \sneia\ arises from at least two different sources: most suffer only a small around of reddening consistent with the properties of interstellar dust in the Milky Way, while for the more highly-extinguished objects the reddening often appears to be produced by dust characterized by an unusually low value of $R_V$.  This may help to explain the rather puzzling result found by many groups that, when $R_V$ is treated as a free parameter in fitting the \sneia\ Hubble diagram, the preferred value is typically $\sim$1.5--2.2 \citep[e.g., see][]{guy10, burns11,sullivan11} since the more-reddened objects will dominate any such calculation.  Moreover, as emphasized by \citet{freedman09} and \citet{guy10}, when minimizing $\chi^2$ in the Hubble diagram, the best-fit value of $R_V$ depends strongly on the errors in the measured colors.  Hence, the fit of $R_V$ not only requires an unbiased estimate of the color, but also of the uncertainties of this estimate.  Underestimated color uncertainties will bias $R_V$ to lower values.

Before leaving this subject, it is worth mentioning a potential complication.  In examining the colors and absolute magnitudes of a nearby sample of \sneia, \citet{wang09} found evidence for two separate subgroups.  The ``Normal'' SNe display a relatively narrow range of expansion velocities of the \ion{Si}{II} $\lambda$6355 absorption near $B$ maximum of 10,600 $\pm$ 400 \kms, whereas the ``HV'' (high velocity) group have \ion{Si}{II} $\lambda$6355 velocities $>$~11,800 \kms.  \citet{wang09} found that the HV \sneia\ had $(B-V)$ colors that were redder on average by $\sim$0.1~mag than the Normal events, and were characterized by $R_V \sim 1.6$ vs. a value of 2.4 for the Normal ones.  However, in a reinterpretation of essentially the same data, \cite{foley11} found that when the most highly-reddened events ($E(B-V) > 0.35$) were eliminated, both the HV and Normal subsamples were consistent with the same value of $R_V = 2.5$.  \citet{foley11} interpreted the redder colors of the HV \sneia\ as being produced by increased line blanketing at blue wavelengths due to the higher expansion velocities.  Fortunately, the HV SNe comprise a relatively small percentage (10--20\%) of the sample observed by the CSP, and the luminosity-color dependence in the NIR is virtually non-existent \citep[see Fig.~14 of][]{burns11}, but these results serve as a caution to assumptions of uniform intrinsic colors for all \sneia.

\subsection{Absolute Magnitudes}
\label{sec:absmags}

The absolute magnitudes of \sneia\ are strongly correlated with decline rate at optical wavelengths \citep{phillips93,hamuy96a,phillips99}.  Although this relationship has allowed \sneia\ to be used as precision distance indicators out to redshifts $z > 1$, the fact that the slope of this relationship becomes progressively shallower at redder wavelengths, led \citet{phillips93} to speculate that ``it might be more fruitful to concentrate observations in the $I$ band, or in the near-infrared, where the intrinsic dispersion in peak brightness appears to be smaller''.  In their pioneering study, \citet{elias85} had already found evidence from a small number of \sneia\ that the intrinsic dispersion in absolute magnitudes in the NIR was as little as $\sim$0.1--0.2~mag.  This was confirmed by \citet{meikle00}, but again from a small sample of only eight SNe, four of which were in the \citet{elias85} sample. A few years later the subject was revisited by \citet{krisciunas04a} who, from a larger sample of 16 \sneia, found dispersions in absolute magnitude of 0.12--0.18~mag in $JHK$, with no evidence for a significant decline rate dependence.  Although it is possible to obtain a similar scatter from the best data sets in $B$ and $V$, it is only achievable after applying significant corrections for {\em both} decline rate and color.  As there is no guarantee that such corrections are the same for all \sneia\ and at all look-back times --- indeed, there is already evidence that they may be dependent on the properties of the host galaxy \citep{hicken09,kelly10,sullivan10,lampeitl10,sullivan11} --- there would seem to be clear advantages to working in the NIR.

In fact, as higher-precision NIR light curves of more and more \sneia\ become available, some evidence for absolute magnitude differences in the NIR are being found.  From a study of 30 \sneia, including nine fast decliners (\dm~$> 1.6$), \citet{krisciunas09} found evidence for a bimodal distribution in the NIR absolute magnitudes at maximum light.  In particular, the fast-declining SNe that peaked in the NIR after the epoch of $B$ maximum were found to be sub-luminous in $JHK$, whereas those that peaked in the NIR before $B$ maximum, were found to have absolute magnitudes indistinguishable from those of normal slower-declining \sneia.  Recently, \citet{kattner11} used a sample of 27 uniformly-observed \sneia\ from the CSP to study the luminosity-decline rate relationship in the NIR.  All 27 SNe had pre-maximum photometry in the optical, and 13 were caught before the NIR maxima.  \citet{kattner11} confirmed the apparent bimodal distribution of absolute magnitudes for fast-declining \sneia.  Moreover, the quality of the data allowed the detection of a weak dependence of luminosity on decline rate for normal \sneia\ at the 2--3$\sigma$ level in the $J$ and $H$ bands.  The dependence in $Y$ was weaker and only detectable at $\sim 1 \sigma$.  This study suggests that applying a correction for decline-rate to \sneia\ luminosities is likely to be beneficial for distance determinations in the $J$ and $H$ bands employing well-sampled, high S/N observations, although the gains are relatively small.  The $Y$ band appears to hold significant promise for future cosmological studies as it offers an optimal combination of signal-to-noise, low sensitivity to dust extinction, and an essentially negligible dependence of absolute magnitude on decline rate.

\subsection{Hubble Diagram}
\label{sec:hubble}

Although \sneia\ are fascinating astrophysical objects on their own, much of the intense interest focussed on them during the last two decades has been motivated by their use as cosmological distance indicators.  The discovery and empirical calibration of the luminosity-decline rate and luminosity-color relationships have made these objects powerful {\em standardizable} candles at optical wavelengths.  Yet there is a growing consensus that their ultimate advantage is in the NIR where, for most intents and purposes, they are truly {\em standard} candles.

\begin{figure}[h]
\begin{center}
\includegraphics[scale=.44, angle=0]{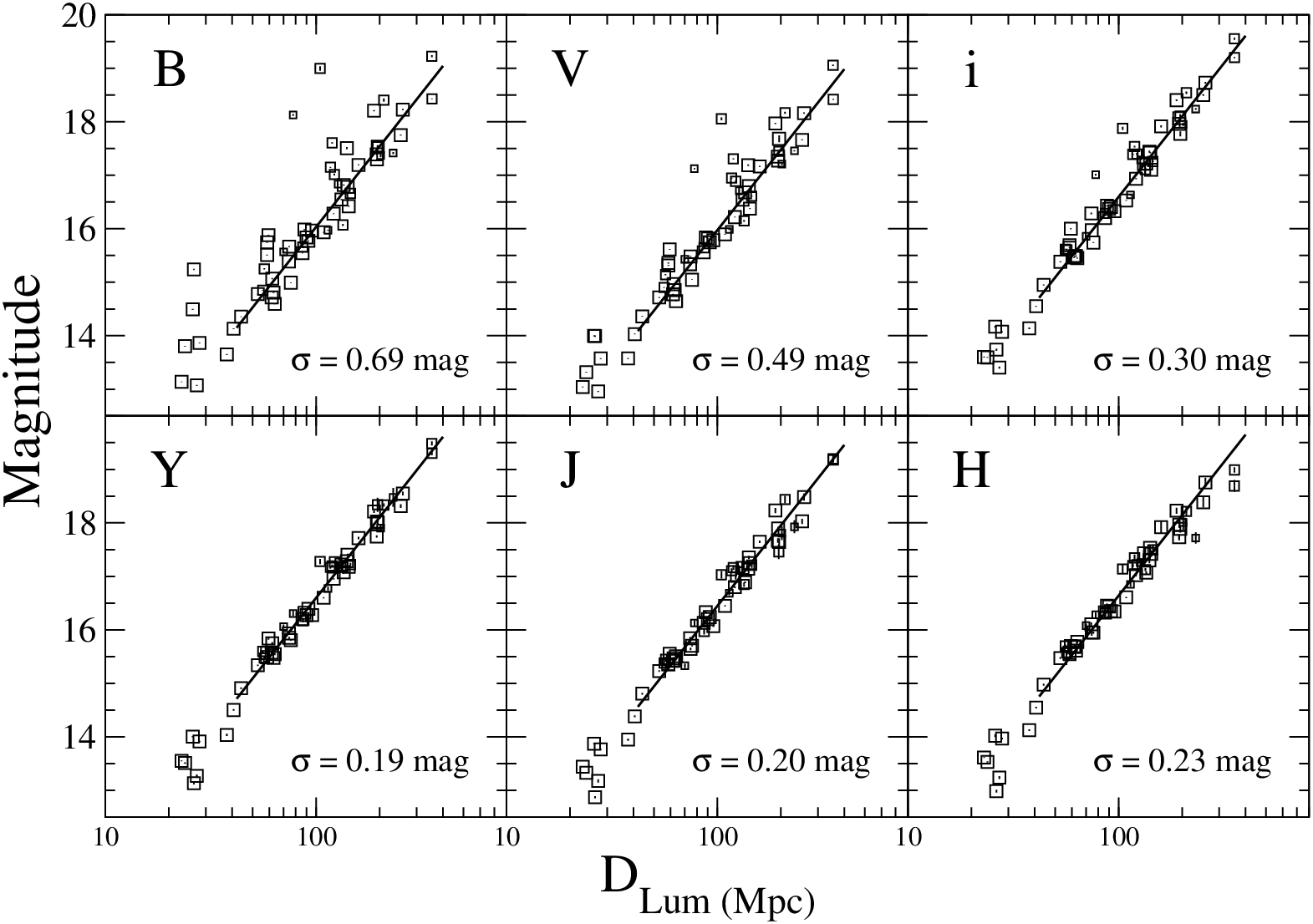}
\caption{Hubble diagrams in $BViYJH$ for 58 \sneia\ observed by the CSP.  The peak magnitudes have been corrected only for Galactic dust extinction.  The best-fit lines are shown.}\label{fig:Hubble_diag}
\end{center}
\end{figure}

To illustrate this, CSP observations in $BViYJH$ for 52 well-observed \sneia\ are plotted as Hubble diagrams in Figure~\ref{fig:Hubble_diag}.  The maximum light magnitudes in all filters were derived from template fits performed with SNooPy.  These were corrected for Milky Way dust extinction using the reddening maps of \citet{schlegel98} and also $K$-corrected, but no luminosity corrections for either decline rate or color were applied.  Note that the dispersion in the optical filters starts off quite high --- 0.69~mag in $B$ --- but drops rapidly to 0.30~mag in the $i$ band, reflecting the fact that the corrections for both empirical relationships are greatest in the blue, but diminish dramatically at redder wavelengths.  The dispersions of 0.19 and 0.20~mag observed in the $Y$ and $J$ bands are truly impressive considering that the data have only been corrected for Galactic reddening.  The slightly larger dispersion of 0.23~mag found in $H$, while still very good, is a bit higher due to the somewhat lower S/N of the photometry, nearly all of which was obtained with the LCO 1~m Swope telescope.

Since the dependence of peak luminosity on decline rate and color is small in the NIR, correcting for these effects produces a relatively small effect on the dispersions, which decreases to 0.17, 0.16, and 0.20~mag in $YJH$, respectively.  However, the median redshift of the CSP \sneia\ is $z \sim 0.02$.  At this distance, the rms error due to peculiar velocities is 5\% in distance ($\pm0.11$~mag in distance modulus) and therefore is a significant contribution to the observed Hubble diagram dispersion \citep{neill07,folatelli10}.  The CSP is now engaged in a new five-year photometric study of an additional $\sim$100 \sneia\ to push the NIR Hubble diagram out to $z \sim 0.08$ in order to determine their true precision as distance indicators.

\section{Fast-Declining \sneia}
\label{sec:fastsneia}

Although evidence had been presented in 1987 that \sneia\ were not all identical \citep{branch87,phillips87}, the discovery and follow-up of two SNe in 1991 --- SN~1991T and SN~1991bg --- definitively demonstrated that \sneia\ show a significant range of properties.  SN~1991T was a slow-declining, luminous SN whose spectrum was dominated by strong \ion{Fe}{III} absorption leading up to maximum, developing the signature \ion{Si}{II} $\lambda$6355 absorption of \sneia\ only after maximum \citep{filippenko92a,phillips92}.  On the other hand, SN~1991bg was a fast-declining, sub-luminous event that was intrinsically red at maximum and showed a deep trough at $\sim$4200~\AA\ due mostly to \ion{Ti}{II} \citep{filippenko92b,leibundgut93}.  As shown a few years later by \citet{nugent95}, these two SNe represent the extremes in a spectral (or effective temperature) sequence of \sneia\ that is strongly correlated with decline rate and has its origin in the amount of $^{56}$Ni produced in the explosion.

The SN~1991T-like objects constitute only 9\% of all \sneia\ in the local Universe \citep{li10}, and fall on the high-luminosity extension of the luminosity-decline rate relationship where they mix with \sneia\ with more normal  spectral characteristics.  On the other hand, the SN~1991bg-like events, which account for 15-20\% of all local \sneia, form more of a separate subgroup and display a strong preference to occur in elliptical and lenticular host galaxies \citep{hamuy96a,hamuy00,li10,gonzalez10}.  An interesting question is whether the 1991bg-like objects are a physically separate subtype of \sneia\ with different progenitors and/or explosion mechanisms \citep[e.g., see][]{pakmor10}, or whether they are the smooth extension of ``normal'' \sneia\ to the smallest $^{56}$Ni masses.  In the latter case, it might be expected that there would be \sneia\ with intermediate properties, the SN~1986G-like events being the most likely candidates \citep{garnavich04}.

\begin{figure}[h]
\begin{center}
\includegraphics[scale=.5, angle=0]{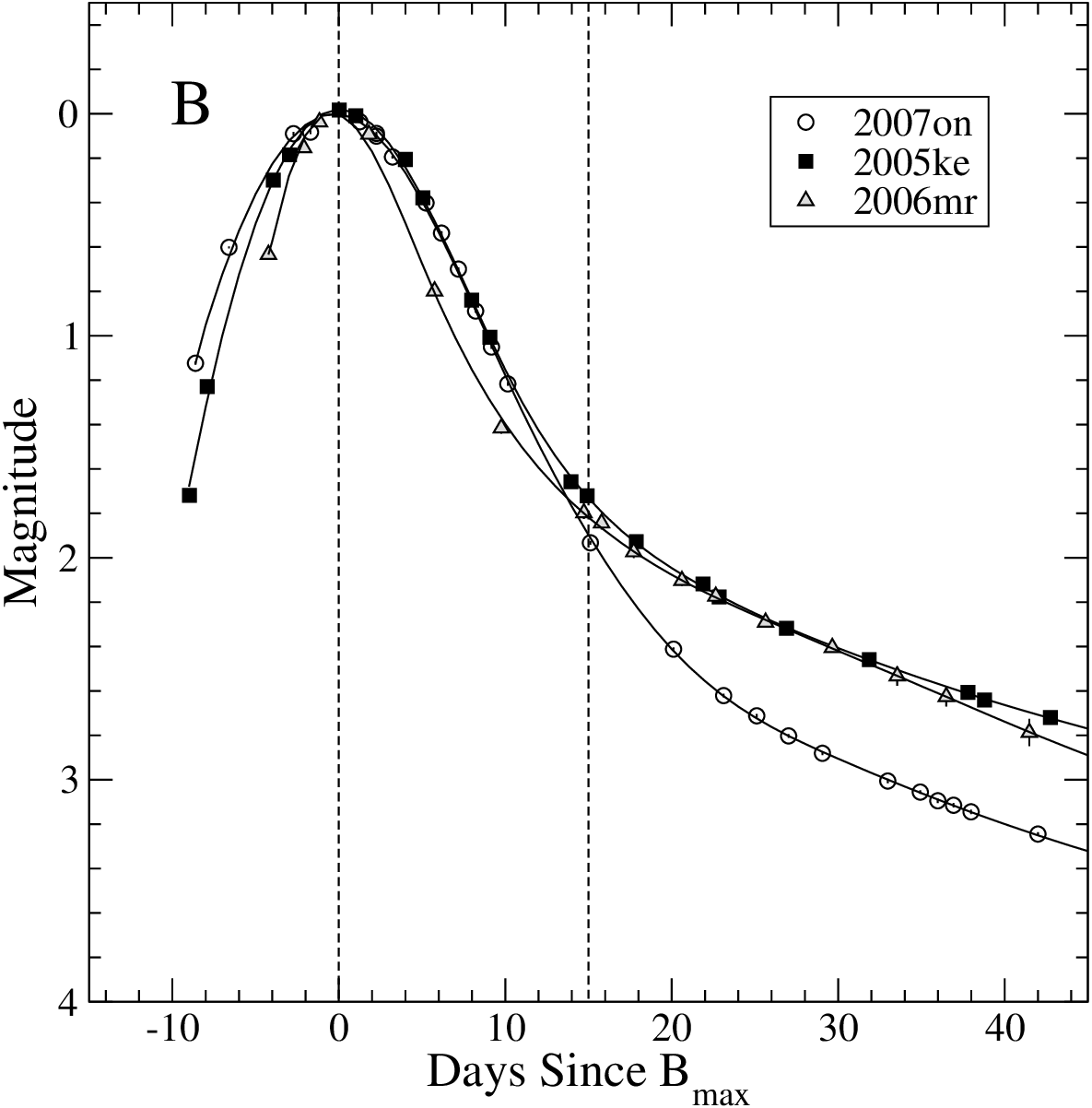}
\caption{$B$ light curves of three fast-declining \sneia\ observed by the CSP.}\label{fig:B_fast}
\end{center}
\end{figure}

Figure~\ref{fig:B_fast} displays the $B$ light curves of three fast-declining \sneia.  As is seen, the value of \dm\ for all three SNe is $\sim$1.9, yet the overall shapes of the light curves are distinct.  SN~2006mr, which was a 1991bg-like event that appeared in NGC~1316 in the outskirts of the Fornax cluster \citep{stritzinger10}, displays the narrowest of the three light curves.  The shape of the light curve of SN~2005ke in NGC~1371, a member of the Eridanus Group, is very similar to that of SN~2006mr after maximum, but displays a slower pre-maximum rise.  Finally, SN~2007on, a 1986G-like event that appeared in the Fornax cluster member NGC~1404, showed the slowest pre-maximum rise, and also faded the most from maximum before reaching the exponential decline phase.  Clearly at these very fast decline rates, the \dm\ parameter loses its power to discriminate between different light-curve morphologies.

\begin{figure}[h]
\begin{center}
\includegraphics[scale=.5, angle=0]{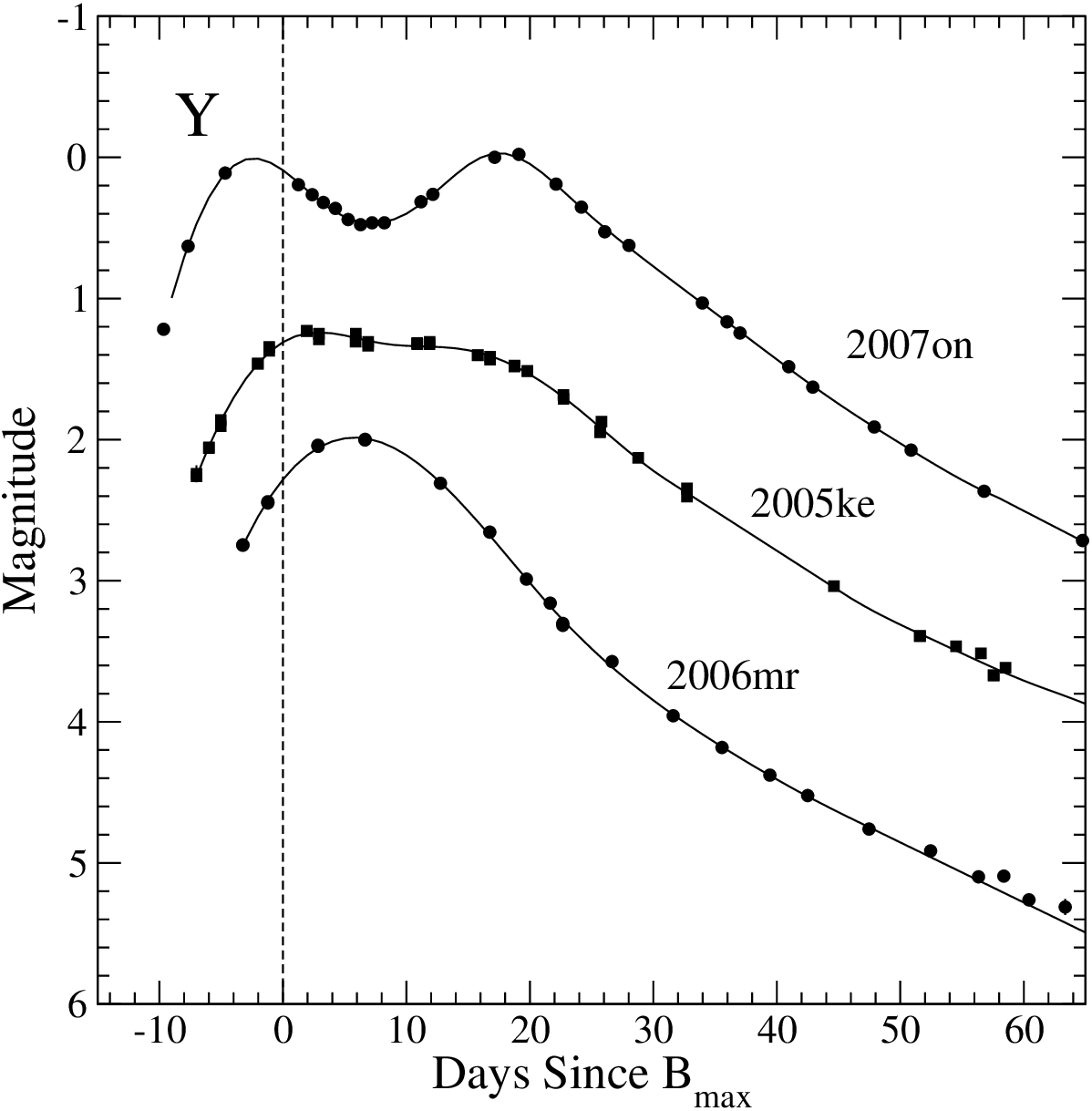}
\caption{Comparison of the $Y$ band light curves of the same three fast-declining \sneia\ shown in Figure~\ref{fig:B_fast}}\label{fig:Y_lcurves_fast}
\end{center}
\end{figure}

The NIR light curves of these three fast-declining \sneia\ are also quite distinct.  This is illustrated in Figure~\ref{fig:Y_lcurves_fast}, which compares the $Y$ band photometry.  The differences between the 1991bg-like SN~2006mr and the 1986G-like SN~2007on are striking.  The primary maximum of the latter peaked a few days before the epoch of $B$ maximum, whereas the light curve of SN~2006mr peaked nearly a week {\em after} maximum.  Also, SN~2007on displayed a clear secondary maximum which is missing in the light curve of SN~2006mr.  These are the distinguishing features of the NIR light curves of 1986G-like and 1991bg-like events as documented by \citet{krisciunas09}.  Interestingly, the $Y$ light curve of SN~2005ke appears to be intermediate between those of the other two SNe; while primary maximum is reached just after the time of $B$ maximum, there is clear evidence of a weak secondary maximum.

\begin{figure}[h]
\begin{center}
\includegraphics[scale=.5, angle=0]{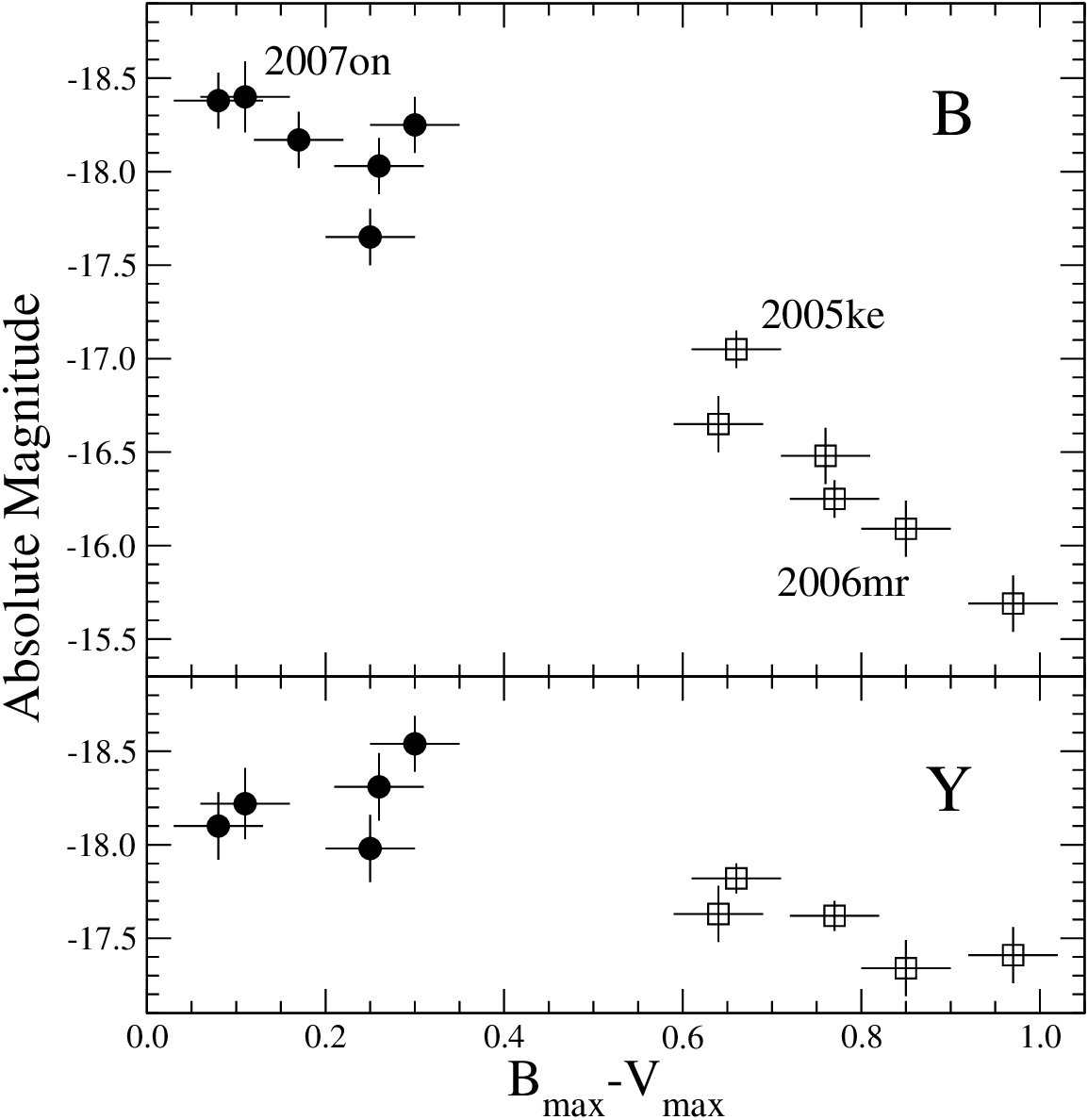}
\caption{Plot of the $(B_{max}-V_{max})$ pseudocolor vs. the $B$- and $Y$-band absolute magnitudes for 12 fast-declining \sneia\ observed by the CSP.  The SNe are divided into 1986G-like (filled circles) and 1991bg-like (open squares) events based on the morphology of the $iYJH$ light curves \citep{krisciunas09}.  All the SNe appeared in early-type hosts and are assumed to have suffered little or no host dust extinction.}\label{fig:B-V_MB}
\end{center}
\end{figure}

Color appears to be a better discriminator than decline rate for the fast-declining \sneia\ \citep{gonzalez10}.  This is shown in Figure~\ref{fig:B-V_MB} where the $(B_{max}-V_{max})$ pseudocolors vs. the $B$ and $Y$ band absolute magnitudes for 12 fast-declining \sneia\ observed by the CSP are plotted.  This sample is divided into 1986G-like and 1991bg-like SNe based on the timing of the primary maximum in $iYJH$ and/or the presence of a secondary maximum \citep{krisciunas09}.  Note that the range of absolute magnitudes is much larger in $B$ than in $Y$.  This is almost certainly due to the strong, temperature-sensitive line blanketing in $B$.  The positions of SNe~2006mr, 2007on, and 2005ke are indicated in Figure~\ref{fig:B-V_MB}.  Again, SN~2005ke is seen to be intermediate in its properties, although closer to those of SN~2006mr.  Interestingly, \citet{krisciunas09} found an apparent correlation between the absolute $J$ magnitude of fast-declining events and the timing of $J$ maximum with respect to the epoch of $B$ maximum, suggesting a smoothly-varying sequence encompassing the 1986G-like and 1991bg-like events.  

\begin{figure}[h]
\begin{center}
\includegraphics[scale=.5, angle=0]{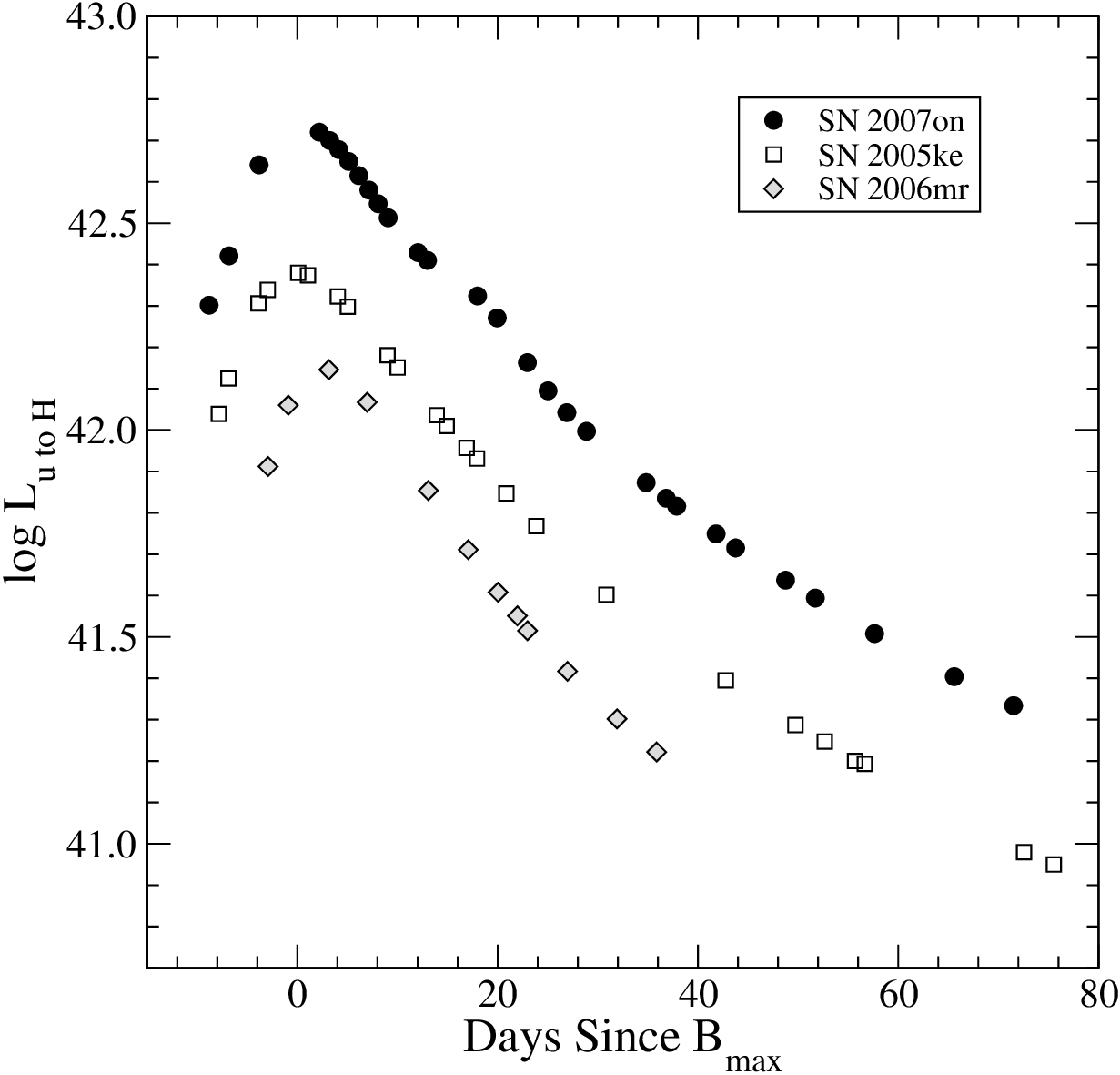}
\caption{Pseudo bolometric light curves of SNe~2006mr, 2007on, and 2005ke.  The luminosity from the $u$ to $H$ photometry obtained by the CSP has been integrated assuming zero host galaxy extinction in all three cases.  The distance moduli assumed for the host galaxies are based on the SBF measurements  of \citet{tonry01}, \citet{jensen03}, and \citet{cantiello07}}\label{fig:uvoir_bol}
\end{center}
\end{figure}

The evolution of the integrated $u$ to $H$ luminosities of SNe~2006mr, 2007on, and 2005ke is plotted in Figure~\ref{fig:uvoir_bol}.  Clear differences in these pseudo-bolometric light curves are observed for all three SNe, not only in the peak luminosity but also in the overall shape of the light curve and the timing of maximum.  Assuming that the peak luminosity is directly proportional to the $^{56}$Ni mass \citep{arnett82}, Figure~\ref{fig:uvoir_bol} implies a factor of $\sim$4 difference in $^{56}$Ni mass between SNe~2007on and 2006mr.

More observations are required before it is possible to give a definitive answer to the question of whether  the 1991bg-like and 1986G-like SNe share similar origins, although the existence of objects such as SN~2005ke that seem to bridge the gap between the two subgroups is certainly suggestive that such a link may exist.

\section{Peculiar \sneia}
\label{sec:pecsneia}

To this point, this review has dealt only with what are commonly deemed ``typical'' \sneia\ which, in this case, include the luminous, slow-declining 1991T-like events as well as the sub-luminous, fast-declining 1991bg-like objects.  Although the vast majority of \sneia\ meet this definition, there are a few members of the class that stick out as truly peculiar.  In this section, what is known of the NIR photometric properties of these objects is briefly reviewed.

\subsection{SN~2002cx-like}
\label{sec:2002cx}

Although the overwhelming majority of \sneia\ obey the luminosity-decline rate relationship and display a remarkably uniform spectral evolution, a few do not.  Such was the case of SN~2002cx, which \citet{li03} called ``the most peculiar known Type~Ia supernova''.  Despite being characterized by a normal initial decline rate of \dm\ $= 1.3$, SN~2002cx displayed a number of peculiar properties including a high-ionization 1991T-like spectrum at maximum light dominated by Fe-group elements, and expansion velocities approximately half those of ordinary \sneia.  The peak absolute magnitudes in $B$ and $V$ were nearly 2~mag fainter than a normal \snia\ of the same decline rate, and the $I$ band light curve displayed a broad primary maximum completely lacking a secondary maximum.

Since the discovery of SN~2002cx, a number of similar SNe have been observed.  Although it has been suggested that the peak luminosities of these objects are highly correlated with both light-curve decline rate and expansion velocities \citep{mcclelland10}, a counter example to the trend was recently identified \citep{narayan11}.

According to \citet{li10}, 2002cx-like events account for $\sim$5\% of all \sneia\ in the local universe.  The first to be observed photometrically in the NIR was SN~2005hk \citep{phillips07}.  Remarkably, the $iYJH$ curves all showed a single broad maximum which, in $Y$ and $H$, did not peak until 15~days after the epoch of $B$ maximum.  \citet{kasen06} has emphasized that the strength of the secondary maximum in the NIR is an excellent diagnostic of the degree of $^{56}$Ni mixing in the ejecta of \sneia.  The double-peaked structure observed in the NIR light curves of typical \sneia\ is a direct sign of the concentration of Fe-peak elements in the central regions, whereas the lack of a secondary maximum is indicative of strong mixing.  This and a number of the other properties of 2002cx-like SNe are most readily explained by the pure deflagration of a white dwarf, although late-time spectroscopy of SN~2002cx showed an unexpectedly high mass and density at low velocity which is not predicted by such a model \citep{jha06}.  Hence, the nature of these objects is still an open question, as is their relationship to normal \sneia.

\subsection{Super-Chandra \sneia}
\label{sec:superchandra}

During the course of the CFHT SN Legacy Survey, a high-redshift ($z = 0.244$) \snia\ was discovered 
with a normal looking spectrum, but with an exceptionally high luminosity ($M_V \sim -20.0$) for its decline rate and unusually low \ion{Si}{II} $\lambda$6355 expansion velocites \citep{howell06}.  The implication was that this object, SN~2003fg (a.k.a. SNLS-03D3bb), had a $^{56}$Ni mass of $1.3 M_\odot$. Such a large $^{56}$Ni mass would seem to be inconsistent with the progenitor having been a Chandrasekhar mass ($1.4 M_\odot$) white dwarf.  The low expansion velocities were also suggestive of a massive progenitor.  In the following years, a few more such SNe have been identified, two of which --- SNe~2007if \citep{scalzo10} and 2009dc \citep[see][and references therein]{taubenberger11} --- were also exceptionally luminous and showed unusually strong absorption features of \ion{C}{II} leading up to maximum light. These objects, which seem to occur preferentially in a low-metallicity environment \citep{childress11,khan11}, have become collectively known as ``super-Chandra \sneia'' --- although it must be said that Chandrasekhar-mass progenitor models have not yet been entirely ruled out.

NIR photometry was obtained of both SNe~2007if and 2009dc by several different groups \citep{scalzo10,yamanaka09,taubenberger11,stritzinger11}.  Both objects had slow decline rates (\dm $\sim$ 0.7), and in both the secondary maximum in the NIR was unusually bright, giving rise to a broad plateau or shoulder.  For SN~2009dc, this phase continued for at least 50 days past maximum in the $J$ band.  Although the merger of two fairly massive white dwarfs to form a super-Chandrasekhar-mass white dwarf has been the most favored model to date, it has been argued that the off-center explosion of a Chandrasekhar-mass white dwarf could also explain the large bolometric luminosity \citep{hillebrandt07}.  However, spectropolarimetry of SN 2009dc indicated that the explosion was quite spherically symmetric \citep{tanaka10}, which appears inconsistent with this idea.  In any case, no model (super-Chandrasekhar or otherwise) has yet successfully explained the observed properties of this apparently rare class of \sneia.

\subsection{SN~2006bt-like}
\label{sec:2006bt}

For those interested in using \sneia\ as cosmological distance indicators, perhaps the most troubling of the peculiar objects are the 2006bt-like SNe.  The prototype of this subclass was observed in detail by \cite{foley10}.  While its decline rate (\dm $= 1.1$) was similar to that of a normal SN~Ia, SN~2006bt displayed intrinsically-red colors and optical spectroscopic properties that were more like those of fast-declining, low-luminosity events.  In addition, the $i$-band light curve showed only a weak secondary maximum.  Although the peak absolute magnitude in $V$ of SN~2006bt was within the dispersion of normal values for \sneia\ with a decline rate of \dm $= 1.1$, the intrinsically-red color evolution of the SN caused standard light curve fitting programs to significantly overestimate the dust reddening.  From Monte Carlo simulations, \citet{foley10} found that these reddening overestimates would get larger as a function of redshift, potentially increasing the Hubble diagram scatter as well as introducing a systematic bias.

\begin{figure}[h]
\begin{center}
\includegraphics[scale=.5, angle=0]{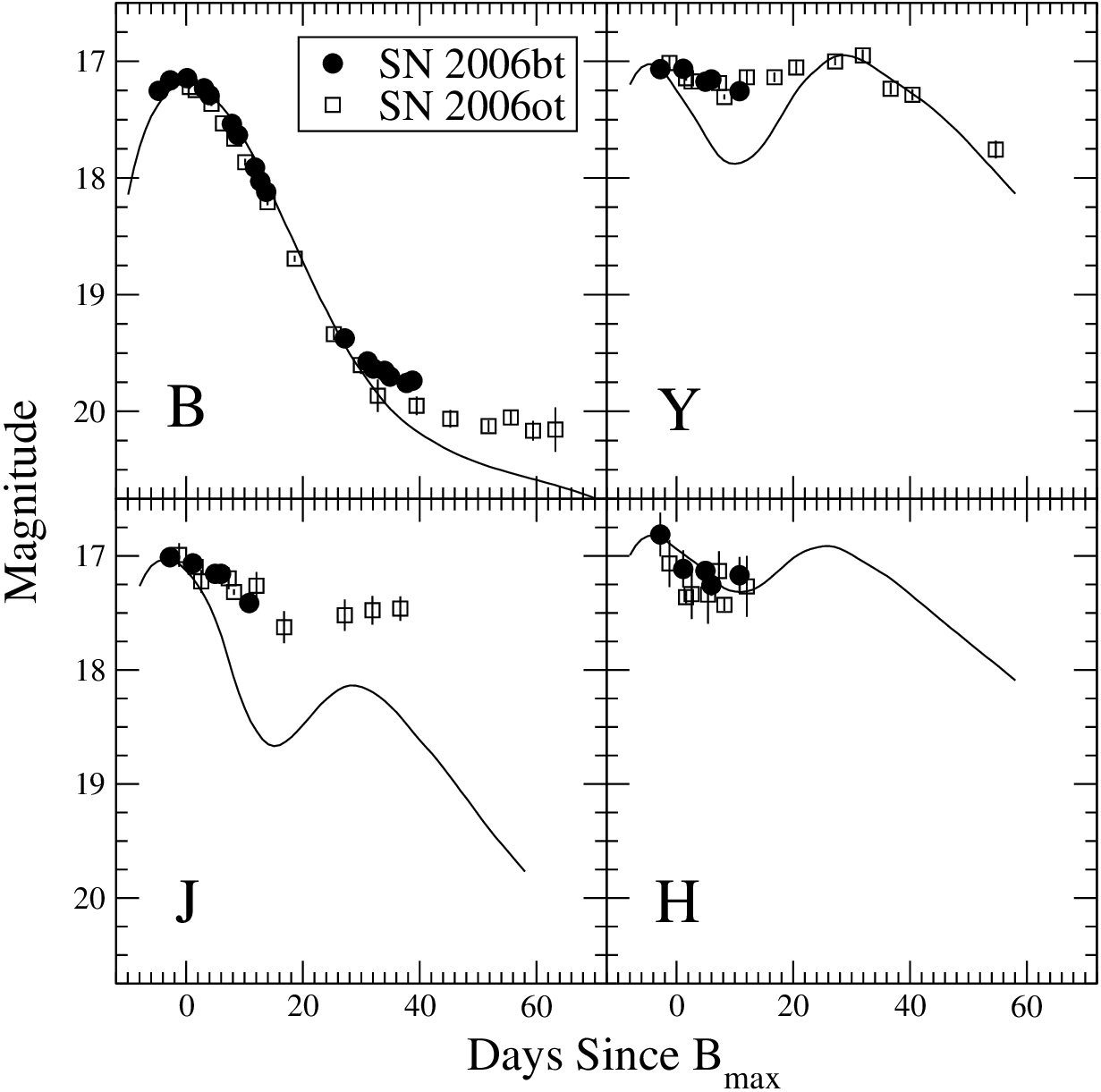}
\caption{$BYJH$ light curves of the peculiar \snia\ 2006bt and the morphologically-similar SN~2006ot.  Template curves of a typical \dm\ $= 1.0$ event are represented by the curved lines.}\label{fig:NIR_06bt_06ot}
\end{center}
\end{figure}

Optical and NIR light curves of SN~2006bt were obtained by the CSP \citep{stritzinger11}, with the $BYJH$ observations reproduced in Figure~\ref{fig:NIR_06bt_06ot}.  For reference, the SNooPY templates of a normal (\dm $= 1.0$) \snia\ are included as solid lines.  In the three NIR filters, broad, slow-declining light curves are observed with the primary and secondary maxima merged to form a plateau-like feature.  This morphology is not unlike that observed for super-Chandra \sneia\ in the same filters, although the peak luminosities are significantly lower for SN~2006bt.  Interestingly, a second object observed by the CSP, SN~2006ot, appears to be closely related to SN~2006bt \citep{stritzinger11}.  The photometry for this object is included in Figure~\ref{fig:NIR_06bt_06ot}, and shows that the two SNe were quite similar.  This similarity also extended to the peak absolute magnitudes, which were the same to within $\sim0.1$~mag.  Spectroscopically, however, SN~2006ot showed clear differences with respect to SN~2006bt, most strikingly in the greater strength and width of the \ion{Si}{II}~$\lambda$6355 line, and its higher expansion velocity.  Interestingly, the hosts of SN~2006bt and SN~2006ot were an S0/a galaxy and an Sa, respectively, implying a relatively
old progenitor \citep{foley10}.

With only two 2006bt-like objects in a sample of 85 \sneia\ observed by the CSP, it appears that these peculiar events are not particularly common.  However, they will be difficult to recognize unless good quality photometry is obtained in the $i$, $Y$, or $J$ bands.   Clearly it is desirable to identify more of these peculiar \sneia\ in order to better document their properties and begin to deduce their relationship to typical \sneia.

\section{Conclusions}
\label{sec:conclu}

Thirty years after the pioneering work of \citet{elias81}, NIR coverage of nearby \sneia\ has become relatively commonplace.  Certainly this change has been helped along by the availability of modern NIR arrays, which have immensely improved both the sensitivity and areal coverage on the sky (important for accurate background subtraction).  These technological improvements have allowed important NIR surveys of \sneia\ such as PARITEL and CSP to be carried out with small (1.0-1.3~m) telescopes.  These observations have revealed the full variety of properties of the NIR light curves of \sneia, from the slowest, most-luminous events to the fastest declining, sub-luminous objects.  In general, an orderly progression of light-curve shapes is observed over this range, although variations exist in the strength of the secondary maximum at any particular decline rate.

NIR photometry has more than fulfilled the promise of improving the precision of \sneia\ as cosmological distance indicators that was originally foreseen in the modest Hubble diagram plotted by \citet{elias85}.  The NIR light curves published for more than 120 nearby \sneia\ during the last ten years have unequivocally demonstrated that these objects are essentially perfect standard candles at these wavelengths, with only a very slight dependence of luminosity on decline rate and color.  In order to build a definitive nearby reference sample, the challenge is not only to observe more SNe, but also to push to higher redshifts since peculiar velocities are the dominant source of the dispersion in the NIR \sneia\ Hubble diagrams published to date.

In order for \sneia\ to be effective in constraining the dark energy equation of state in next-generation experiments such as WFIRST, the evolution of the empirically-corrected luminosity must be less than 1-2\% out to $z < 1.5$ \citep{albrecht06}.  NIR observations appear to be the best (perhaps the only) way to meet such a requirement.  At an effective wavelength of 1.03~$\mu$m, the $Y$ band seems extremely well suited for pushing to high redshift since it offers the best combination of shortest wavelength, achievable signal-to-noise, and insensitivity to luminosity corrections for decline rate and color.  The NIR also provides an effective way to identify certain peculiar \sneia\ whose optical light curves are otherwise normal in appearance.

This review has emphasized the advances made in our understanding of the photometric properties of \sneia, but the NIR spectroscopic properties of these objects are still relatively poorly documented.  While important progress in this area was made recently with the publication of 41 spectra of 29 \sneia\ by \citet{marion09}, aside from the normal SNe~1999ee \citep{hamuy02} and 2003du \citep{stanishev07} and the fast-decliner SN~1999by \citep{hoeflich02}, there are only a few events for which multi-epoch spectra have been obtained.  Such data offer considerable promise for constraining the radial distribution of the elements synthesized in the explosion, particularly since the optical depth in the NIR drops rapidly after maximum, exposing a greater amount of the SN ejecta.  The possible existence of metallicity and luminosity spectroscopic indicators in the NIR is an area that is also relatively unexplored.  Finally, a better understanding of the NIR spectral evolution of \sneia\ is vital for determining precise $K$ corrections for future cosmology studies.  With the growing availability of fast NIR spectrographs on 4--8~m telescopes, this is an area where we can expect much progress in coming years.

\section*{Acknowledgments} 

The author wishes to acknowledge the Oskar Klein Centre for Cosmoparticle Physics at the University of Stockholm and, in particular, Ariel Goobar and Maximilian Stritzinger, for their hospitality and support during a 3-week visit in May 2011, when work on this review paper began.  Special thanks are also in order to Chris Burns, Eric Hsiao, and Maximilian Stritzinger for their valuable comments on an early draft of this paper.  Many of the NIR observations discussed in this review were obtained by the CSP, which has been generously supported by the National Science Foundation under grants ASTÐ0306969, ASTÐ0607438, and ASTÐ1008343.  The CSP began in 2004, and the fruits that it is now bearing have resulted from the dedication and effort of a large number of people with whom it has been my pleasure to work.

\begin{thebibliography}{}
\setlength{\itemsep}{0.0mm}

\bibitem[Albrecht et al.(2006)]{albrecht06} Albrecht, A., et al. 2006, ``Report of the Dark Energy Task Force'', (arXiv:astroph/0609591)
\bibitem[Arnett(1982)]{arnett82} Arnett, W.~D. 1982, Astrophysical Journal, 253, 785
\bibitem[Axelrod(1980)]{axelrod80} Axelrod, T.~S. 1980, Ph.D. Thesis, University of California, Santa Cruz 
\bibitem[Branch(1987)]{branch87} Branch, D. 1987, Astrophysical Journal, 316, L81
\bibitem[Burns et al.(2011)]{burns11} Burns, C., et~al. 2011, Astronomical Journal, 141, 19
\bibitem[Cantiello et al.(2007)]{cantiello07} Cantiello, M., Blakeslee, J., Raimondo, G., Brocato, E., Capaccioli, M. 2007, Astrophysical Journal, 668, 130
\bibitem[Cardelli, Clayton, \& Mathis(1989)Cardelli et al.]{cardelli89} Cardelli, J.~A., Clayton, G.~C., Mathis, J~.S. 1989, Astrophysical Journal, 345, 245 
\bibitem[Childress et al.(2011)]{childress11} Childress, M., et al. 2011, Astrophysical Journal, 733, 3
\bibitem[Chotard et al.(2011)]{chotard11} Chotard, N., et~al. 2011, Astronomy and Astrophysics, 529, L4
\bibitem[Cristiani et al.(1992)]{cristiani92} Cristiani, S., et~al. 1992, Astronomy and Astrophysics, 259, 63
\bibitem[Contreras et al.(2010)]{contreras10} Contreras, C., et~al. 2010, Astronomical Journal, 139, 519
\bibitem[Elias et al.(1981)]{elias81} Elias, J.~H., Frogel, J.~A., Hackwell, J.~A., Persson, S.~E. 1981, Astrophysical Journal, 251, L13
\bibitem[Elias \& Frogel(1983)]{elias83} Elias, J.~H., Frogel, J.~A. 1983, Astrophysical Journal, 268, 718
\bibitem[Elias et al.(1985)]{elias85} Elias, J.~H., Matthews, K., Neugebauer, G., Persson, S.~E. 1985, Astrophysical Journal, 296, 379
\bibitem[Filippenko et al.(1992a)]{filippenko92a} Filippenko, A.~V., et al. 1992a, Astrophysical Journal, 348, L15
\bibitem[Filippenko et al.(1992b)]{filippenko92b} Filippenko, A.~V., et al. 1992b, Astronomical Journal, 104, 1543
\bibitem[Folatelli et al.(2010)]{folatelli10} Folatelli,~G., et al. 2010, Astronomical Journal, 139, 120
\bibitem[Foley et al.(2010)]{foley10} Foley, R.~J., et al. 2010, Astrophysical Journal, 708, 1748
\bibitem[Foley \& Kasen(2011)]{foley11} Foley, R.~J., Kasen, D. 2010, Astrophysical Journal, 729, 55
\bibitem[Freedman et al.(2009)]{freedman09} Freedman, W.~L., et al. 2009, Astrophysical Journal, 704, 1036
\bibitem[Frogel et al.(1987)]{frogel87} Frogel, J.~A., Gregory, B., Kawara, K., Laney, D., Phillips, M.~M., Terndrup, D., Vrba, F., Whitford, A.~E. 1987, Astrophysical Journal, 315, L129
\bibitem[Garnavich et al.(2004)]{garnavich04} Garnavich, P.~M. 2004, Astrophysical Journal, 613, 1120
\bibitem[Gonz\'{a}lez-Gait\'{a}n et al.(2010)]{gonzalez10} Gonz\'{a}lez-Gait\'{a}n, S. 2010, Astrophysical Journal, 727, 107
\bibitem[Goobar(2008)]{goobar08} Goobar, A. 2008, Astrophysical Journal, 686, L103
\bibitem[Guy et al.(2010)]{guy10} Guy, J., et al. 2010, Astronomy and Astrophysics, 523, 7
\bibitem[Hamuy et al.(1996a)]{hamuy96a} Hamuy,~M., et~al. 1996a, Astronomical Journal, 112, 2391
\bibitem[Hamuy et al.(1996b)]{hamuy96b} Hamuy,~M., et~al. 1996b, Astronomical Journal, 112, 2438
\bibitem[Hamuy et al.(2000)]{hamuy00} Hamuy, M., Trager, S.~C., Pinto, P.~A., Phillips, M.~M., Schommer, R.~A., Ivanov, V., Suntzeff, N.~B. 2000, Astronomical Journal, 120, 1479
\bibitem[Hamuy et al.(2002)]{hamuy02} Hamuy, M., et al. 2002, Astronomical Journal, 124, 417
\bibitem[Hamuy et al.(2006)]{hamuy06} Hamuy, M., et al. 2006, Publications of the Astronomical Society of the Pacific, 118, 2
\bibitem[Hernandez et al.(2000)]{hernandez00} Hernandez, M., et~al. 2000, Monthly Notices of the Royal Astronomical Society, 319, 223
\bibitem[Hicken et al.(2009)]{hicken09} Hicken, M., Wood-Vasey, W.~M., Blondin, S., Challis, P., Jha, S., Kelly, P.~L., Rest, A., Kirshner, R.~P. 2009, Astrophysical Journal, 700, 1097
\bibitem[Hillebrandt, Sim, \& R\"{o}pke(2007)]{hillebrandt07} Hillebrandt, W., Sim, S.~A.,~R\"{o}pke, F.~K. 2007, Astronomy and Astrophysics, 465, L17
\bibitem[H\"{o}flich, Khokhlov, \& Wheeler(1995)H\"{o}flich et al.]{hoeflich95} H\"{o}flich, P., Khokhlov, A.~M., Wheeler, J.~C. 1995, Astrophysical Journal, 444, 831
\bibitem[H\"{o}flich et al.(2002)]{hoeflich02} H\"{o}flich, P., Gerardy, C., Fesen, R., Sakai, S. 2002, Astrophysical Journal, 568, 791
\bibitem[Howell et al.(2006)]{howell06} Howell, D.~A., et al. 2006, Nature, 443, 308
\bibitem[Hoyle \& Fowler(1960)]{hoyle60} Hoyle, F., Fowler, W. A. 1960, Astrophysical Journal, 132, 565
\bibitem[Jha et al.(1999)]{jha99} Jha, S., et~al. 1999, Astrophysical Journal Supplements, 125, 73
\bibitem[Jha et al.(2006)]{jha06} Jha, S., Branch, D., Chornock, R., Foley, R.~J., Li, W., Swift, B.~J., Casebeer, D., Filippenko, A.~V. 2006, Astronomical Journal, 132, 189
\bibitem[Jensen et al.(2003)]{jensen03} Jensen, J.~B., Tonry, J.~L., Barris, B.~J., Thompson, R.~I., Liu, M.~C., Rieke, M.~J., Ajhar, E.~A., Blakeslee, J.~P. 2003, Astrophysical Journal, 583, 712
\bibitem[Jha, Riess, \& Kirshner(2007)Jha et al.]{jha07} Jha, S., Riess, A.~G., Kirshner, R.~P. 2007, Astrophysical Journal, 659, 122
\bibitem[Kasen(2006)]{kasen06} Kasen, D. 2006, Astrophysical Journal, 649, 939
\bibitem[Kattner et al.(2011)]{kattner11} Kattner, S., et al. 2011, Publications of the Astronomical Society of the Pacific, submitted
\bibitem[Kelly et al.(2010)]{kelly10} Kelly, P.~L., Hicken, M., Burke, D.~L., Mandel, K.~S., Kirshner, R.~P. 2010, Astrophysical Journal, 715, 743
\bibitem[Kessler et al.(2009)]{kessler09} Kessler, R., et al. 2009, Astrophysical Journal Supplements, 182, 32
\bibitem[Khan et al.(2011)]{khan11} Khan, R., Stanek, K.~Z., Stoll, R., Prieto, J.~L. 2011, Astrophysical Journal, 737, L24
\bibitem[Kirshner et al.(1973)]{kirshner73} Kirshner, R.~P., Willner, S.~P., Becklin, E.~E., Neugebauer, G., Oke, J.~B. 1973, Astrophysical Journal, 180, L97
\bibitem[Krisciunas et al.(2000)]{krisciunas00} Krisciunas, K., Hastings, N.~C., Loomis, K., McMillan, R., Rest, A., Riess, A.~G., Stubbs, C. 2000, Astrophysical Journal, 539, 658
\bibitem[Krisciunas et al.(2001)]{krisciunas01} Krisciunas, K., et~al. 2001, Astronomical Journal, 122, 1616
\bibitem[Krisciunas et al.(2003)]{krisciunas03} Krisciunas, K., et al. 2003, Astronomical Journal, 125, 166 
\bibitem[Krisciunas, Phillips, \& Suntzeff(2004a)Krisciunas et al.]{krisciunas04a} Krisciunas, K., Phillips, M.~M., Suntzeff, N.~B. 2004a, Astrophysical Journal, 602, L81
\bibitem[Krisciunas et al.(2004b)]{krisciunas04b} Krisciunas, K., et al. 2004b, Astronomical Journal, 127, 1664
\bibitem[Krisciunas et al.(2004c)]{krisciunas04c} Krisciunas, K., et al. 2004c, Astronomical Journal, 128, 3034
\bibitem[Krisciunas et al.(2007)]{krisciunas07} Krisciunas,~K., et~al. 2007, Astronomical Journal, 133, 58
\bibitem[Krisciunas et al.(2009)]{krisciunas09} Krisciunas,~K., et~al. 2009, Astronomical Journal, 138, 1584
\bibitem[Lampeitl et al.(2010)]{lampeitl10} Lampeitl, H., et~al. 2010, Astrophysical Journal, 722, 566
\bibitem[Leibundgut et al.(1993)]{leibundgut93} Leibundgut, B., et al. 1992, Astronomical Journal, 105, 301
\bibitem[Leloudas et al.(2009)]{leloudas09} Leloudas, G., et~al. 2009, Astronomy and Astrophysics, 505, 265
\bibitem[Li et al.(2003)]{li03} Li, W., et al. 2003, Publications of the Astronomical Society of the Pacific, 115, 453
\bibitem[Li et al.(2010)]{li10} Li, W., Chornock, R., Leaman, J., Filippenko, A.~V., Poznanski, D., Wang, X., Ganeshalingam, M., Mannucci, F. 2010, Monthly Notices of the Royal Astronomical Society, 412, 1473
\bibitem[Mandel, Narayan, \& Kirshner(2011)Mandel et al.]{mandel11} Mandel, K.~S., Narayan, G., Kirshner, R.~P. 2011, Astrophysical Journal, 731, 120
\bibitem[Marion et al.(2009)]{marion09} Marion, G.~H., et al. 2009, Astronomical Journal, 138, 727
\bibitem[McClelland et al.(2010)]{mcclelland10} McClelland, C.~M. 2010, Astrophysical Journal, 720, 704
\bibitem[Meikle(2000)]{meikle00} Meikle, W.~P.~S. 2000, Monthly Notices of the Royal Astronomical Society, 314, 782
\bibitem[Minkowski(1941)]{minkowski41} Minkowski, R. 1941, Publications of the Astronomical Society of the Pacific, 53, 224
\bibitem[Narayan et al.(2011)]{narayan11} Narayan, G., et al. 2011, Astrophysical Journal, 731, L11
\bibitem[Neill, Hudson, \& Conley(2007)Neill et al.]{neill07} Neill, J.~D., Hudson, M.~J., Conley, A. 2007, Astrophysical Journal, 661, L123
\bibitem[Nugent et al.(1995)]{nugent95} Nugent, P., Phillips, M., Baron, E., Branch, D., Hauschildt, P. 1995, Astrophysical Journal, 455, L147
\bibitem[Nugent et al.(2011)]{nugent11} Nugent, P.~E., Sullivan, M., Bersier, D., Howell, D.~A., Thomas, R., James, P. 2011, Central Bureau for Astronomical Telegrams, Electronic Telegram No. 2792
\bibitem[Pakmor et al.(2010)]{pakmor10} Pakmor, R., Kromer, M., R\"{o}pke, F.~K., Sim, S.~A., Ruiter, A.~J.,
Hillebrandt, W. 2010, Nature, 463, 61
\bibitem[Patat et al.(2007)]{patat07} Patat, F., et al. 2007, Science, 317, 924
\bibitem[Perlmutter(1997)]{perlmutter97} Perlmutter, S. 1997, in Thermonuclear Supernovae, ed. P. Ruiz-Lapuente, R. Canal, \& J. Isern (Dordrecht: Kluwer), 749
\bibitem[Perlmutter et al.(1999)]{perlmutter99} Perlmutter, S., et~al. 1999, Astrophysical Journal, 517, 565
\bibitem[Phillips et al.(1987)]{phillips87} Phillips, M.~M., et~al. 1987, Publications of the Astronomical Society of the Pacific, 99, 592
\bibitem[Phillips et al.(1992)]{phillips92} Phillips, M.~M., et al. 1992, Astronomical Journal, 103, 1632
\bibitem[Phillips(1993)]{phillips93} Phillips, M.~M. 1993, Astrophysical Journal, 413, L105
\bibitem[Phillips et al.(1999)]{phillips99} Phillips, M.~M., Lira, P., Suntzeff, N.~B., Schommer, R.~A., Hamuy, M., \& Maza, J. 1999, Astronomical Journal, 118, 1766 
\bibitem[Phillips et al.(2007)]{phillips07} Phillips, M.~M. 2007, Publications of the Astronomical Society of the Pacific, 119, 360
\bibitem[Pinto \& Eastman(2000)]{pinto00} Pinto, P.~A., Eastman, R.~G. 2000, ApJ, 530, 757
\bibitem[Riess et al.(1998)]{riess98} Riess, A.~G., et~al. 1998, Astronomical Journal, 116, 1009
\bibitem[Riess et al.(2011)]{riess11} Riess, A.~G., et~al. 2011, Astrophysical Journal, 730, 119
\bibitem[Scalzo et al.(2010)]{scalzo10} Scalzo, R.~A., et al. 2010, Astrophysical Journal, 713, 1073
\bibitem[Schlegel, Finkbeiner, \& Davis(1998)Schlegel et al.]{schlegel98} Schlegel, D.~J., Finkbeiner, D.~P., Davis,~M. 1998, Astrophysical Journal, 500, 525
\bibitem[Sollerman et al.(2004)]{sollerman04} Sollerman, J., et~al. 2004, Astronomy and Astrophysics, 428, 555
\bibitem[Spyromilio, Pinto, \& Eastman (1994)Spyromilio et al.]{spyromilio94} Spyromilio, J., Pinto, P.~A., Eastman, R.~G. 1994, Monthly Notices of the Royal Astronomical Society, 266, L17
\bibitem[Spyromilio et al.(2004)]{spyromilio04} Spyromilio, J., Gilmozzi, R., Sollerman, J., Leibundgut, B., Fransson, C., Cuby, J.-G. 2004, Astronomy and Astrophysics, 426, 547
\bibitem[Stanishev et al.(2007)]{stanishev07} Stanishev, V., et~al. 2007, Astronomy and Astrophysics, 469, 645
\bibitem[Stritzinger \& Sollerman(2007)]{stritzinger07} Stritzinger, M, Sollerman, J. 2007, Astronomy and Astrophysics, 470, L1
\bibitem[Stritzinger et al.(2010)]{stritzinger10} Stritzinger, M.~D. 2010, Astronomical Journal, 140, 2036
\bibitem[Stritzinger et al.(2011)]{stritzinger11} Stritzinger, M.~D. 2011, Astronomical Journal, in press
\bibitem[Sullivan et al.(2010)]{sullivan10} Sullivan, M., et al. 2010, Monthly Notices of the Royal Astronomical Society, 406, 782
\bibitem[Sullivan et al.(2011)]{sullivan11} Sullivan, M., et al. 2011, Astrophysical Journal, 737, 102
\bibitem[Tanaka et al.(2010)]{tanaka10} Tanaka, M., et al. 2010, Astrophysical Journal, 714, 1209
\bibitem[Taubenberger et al.(2011)]{taubenberger11} Taubenberger, S. et al. 2011, Monthly Notices of the Royal Astronomical Society, 412, 2735
\bibitem[Tonry et al.(2001)]{tonry01} Tonry, J.~L., Dressler, A., Blakeslee, J.~P., Ajhar, E.~A., Fletcher, A.~B., Luppino, G.~A., Metzger, M.~R., Moore, C.~B. 2001, Astrophysical Journal, 546, 681 
\bibitem[Wang et al.(2009)]{wang09} Wang, X., et al. 2009, Astrophysical Journal, 699, L139
\bibitem[Wood-Vasey et al.(2008)]{wood-vasey08} Wood-Vasey, W.~M., et al. 2008, Astrophysical Journal, 689, 377
\bibitem[Yamanaka et al.(2009)]{yamanaka09} Yamanaka, M., et al. 2009, Astrophysical Journal, 707, L118 
\end{thebibliography}
\end{document}